\documentstyle[epsf,twoside,fleqn,espcrc2]{article}
\title{Gauge invariant field strength correlators and Abelian dominance in QCD}
\author{Nora Brambilla
\address{Institut f\"ur Theoretische Physik, Univ. Wien, Boltzmanngasse 5, A-1090 Vienna, Austria}
\thanks{Marie Curie fellow, TMR contract n. ERBFMBICT961714}        
and Antonio Vairo
\address{Institut f\"ur Hochenergiephysik, \"Oster. Akad. der Wiss., Nikolsdorfergasse 18, 
         A-1050 Vienna, Austria}}

\begin{document}

\begin{abstract}
We establish a relation between the two-point field strength correlator in QCD and the dual field 
propagator of the effective dual Abelian Higgs model describing the infrared behaviour of QCD.
\end{abstract}

\maketitle

\section{INFRARED ABELIAN DOMINANCE}

The QCD infrared dynamics  appears to be well approximated by a (dual) Abelian Higgs
model \cite{polikarpov,bali0}. Via Abelian projection  the QCD gluodynamics is reduced 
to Abelian fields, Abelian monopoles and charged matter fields degrees of freedom. 
Then, lattice simulations (mainly in $SU(2)$ and in Maximal Abelian projection) 
show Abelian dominance and monopole dominance in the long range features of QCD. 
This means that nonperturbative dynamical quantities like the string tension $\sigma$ 
are reproduced (for more than $90\%$) simply by considering the Abelian or the monopole degrees 
of freedom.  Moreover, monopoles are condensed in the confined phase. Electric 
fields and magnetic currents in the presence of static sources are consistent with the dual 
Maxwell and Ginzburg--Landau equations. The penetration length $\displaystyle \lambda= {1/M}$ 
($M=$ dual gluon mass) and the correlation length $\displaystyle \xi={1/M_\phi}$ 
($M_\phi=$ Higgs Mass) are measured. Finally, the effective monopole Lagrangian is  extracted 
from an inverse Montecarlo method and turns out to be equivalent to the Lagrangian 
of the dual Abelian Higgs model. Therefore, the above lattice investigations show that 
1) QCD lattice gluodynamics is {\it equivalent} to the dual Abelian 
Higgs model: the Higgs particles are Abelian monopoles and condense in the confinement phase.
2) $M \simeq M_\phi$  and therefore the QCD vacuum behaves as a dual superconductor 
on the border between type I and type II.

These are the information coming from the lattice. A still open problem is how 
to relate the Higgs and the dual gluon mass with the QCD parameters. In the following we show 
that an ``Abelian effective dominance'' shows up in the cumulant expansion of the 
quasi-static Wilson loop average in the nonperturbative region. This allows us to relate 
the non-local gluon condensate (i.e. the two-point field strength correlator) 
with the dual field propagator of an underlying dual Abelian Higgs model and 
therefore to suggest a connection between QCD and dual parameters.

\section{WILSON LOOP AND FIELD STRENGTH CORRELATORS}

In this section we report evidence of the {\it Gaussian dominance} in the cumulant expansion 
of the quasi-static Wilson loop average. The infrared gluodynamics is contained in the Wilson loop 
which is an order parameter of confinement
$W(\Gamma)\equiv {\rm P}  \exp \{ig \displaystyle \oint_\Gamma dz_\mu A_\mu(z)\}$
and $\Gamma$ is a loop made up by the two quark trajectories plus the endpoint Schwinger strings.
In Euclidean space, using the non-Abelian Stokes theorem, the Wilson loop average 
on the Yang--Mills action,  $\displaystyle \langle ~~\rangle \equiv 
\int {\cal D}A \exp \{i S_{\rm YM}[A]\} $, can be expanded as \cite{DoSi},
\begin{eqnarray}
&& \!\!\!\!\!\!\!\!\!\!\!\!\!\!   \langle W(\Gamma)\rangle = 
 \exp \sum_{n=0}^\infty {(ig)^n \over n!} \int_{S(\Gamma)} dS(1) \cdots  dS(n)
\nonumber \\
& &\!\!\!\!\!\!\!\!\!\!\!  
\times \langle \phi(0,1) F(1)\phi(1,0) \cdots F(n)\phi(n,0) \rangle_{\rm cum}.
\label{cumex}
\end{eqnarray}
where $F(n) \equiv F_{\mu_n \nu_n}(u_n)$ is the field strength. $S(\Gamma)$ denotes a surface 
with contour $\Gamma$ and $dS(n) \equiv dS_{\mu_n \nu_n}(u_n)$. 
The cumulants $\langle ~~ \rangle_{\rm cum}$ are defined as the connected 
part of the average and  $\langle \phi(0,1) F(1) \phi(1,0) \rangle_{\rm cum} 
= \langle \phi(0,1) F(1) \phi(1,0) \rangle$ $=$ $0$. 
$\phi(0,u)= \exp \{ ig \int_0^u dz_\mu A_\mu (z) \}$ is a Schwinger line. 

Let us consider $\langle\!\langle g^2F_{\mu\nu}(x)F_{\lambda\rho}(0)\rangle\!\rangle_\Gamma 
- \langle\!\langle gF_{\mu\nu}(x)\rangle\!\rangle_\Gamma 
\langle\!\langle gF_{\lambda\rho}(0)\rangle\!\rangle_\Gamma$, with $\Gamma$ a rectangular 
Wilson loop and the points $x\equiv({\bf r},t)$ and $0$ belonging to the first and second 
temporal quark line respectively. This object is required to compute the $1/m^2$ corrections to the
static quark-antiquark potential and has been evaluated on the lattice.  The double bracket
$\langle\!\langle ~~ \rangle\!\rangle_\Gamma$ stands for the
average over gauge fields in presence of the Wilson loop $W(\Gamma)$
$\langle\!\langle ~~\rangle\!\rangle_\Gamma \equiv \langle~~ W(\Gamma) \rangle / 
\langle W(\Gamma) \rangle$. Taking into account Eq. (\ref{cumex}), we obtain,
\begin{eqnarray}
 & & \!\!\!\!\!\!\!\!\!\!\!\!\!\!
\langle\!\langle g^2F_{\mu\nu}(x)F_{\lambda\rho}(0)\rangle\!\rangle_\Gamma 
- \langle\!\langle gF_{\mu\nu}(x)\rangle\!\rangle_\Gamma 
  \langle\!\langle gF_{\lambda\rho}(0)\rangle\!\rangle_\Gamma =
\nonumber \\
& & \!\!\!\!\!\!\!\!\!\!\!\!\!\! 
g^2\langle \phi(0,x) F_{\mu\nu}(x) \phi(x,0) F_{\lambda\rho} (0) \rangle  
+ {\cal R}_{\mu\nu\lambda\rho}(x;\Gamma),
\label{FF} 
\end{eqnarray} 
where $\cal R$ contains all contributions coming from cumulants of order higher than two:
\begin{eqnarray}
& & \!\!\!\!\!\!\!\!\!\!\!\!  {\cal R}_{\mu\nu\lambda\rho}(x;\Gamma) =
\nonumber \\ 
& & \!\!\!\!\!\!\!\!\!  \sum_{n=3}^\infty 
{(ig)^n\over n!} \!\!\! \sum_{\{ {\rm all \, perm.} (i,j) \} } 
\int_{S(\Gamma)} \left[ \prod_{k \neq i,j}^n dS_{\mu_k\nu_k}(u_k)\right] 
\nonumber\\
&& \!\!\!\!\!\!\!\!\!\!\!\!  \times  \langle \phi(0,u_1) F_{\mu_1\nu_1}(u_1) \cdots 
F_{\mu\nu}(u_i = x) 
\nonumber\\
&&\!\!\!\!\!\!\!\!\!\!\! \times \cdots F_{\lambda\rho}(u_j=0) \cdots  
F_{\mu_n\nu_n}(u_n)\phi(u_n,0) \rangle_{\rm cum}.
\nonumber
\label{rest}
\end{eqnarray} 

In the so-called Stochastic Vacuum Model \cite{DoSi} it is assumed that for
large distances the bilocal cumulant $\langle \phi(0,x) F_{\mu\nu}(x)
\phi(x,0) F_{\lambda\rho} (0) \rangle$ is the dominant contribution to
the r.h.s. in Eq. (\ref{FF}).  This assumption has been (phenomenologically) successfully 
tested in potential theory and in soft high energy scattering processes. Eq. (\ref{FF}) 
is suitable to obtain a first principles check of this assumption on the lattice. 
The adopted strategy is the following \cite{bali}. Within Eq. (\ref{FF}) Lorentz invariance is 
only violated by the contour $\Gamma$. Therefore, deviations of the l.h.s. of Eq. (\ref{FF}) 
from Lorentz invariance have to be attributed to the higher order cumulants $\cal R$ on the r.h.s. 
We interpret the Lorentz invariant part of the data as $\langle \phi(0,x) F_{\mu\nu}(x)
\phi(x,0) F_{\lambda\rho} (0) \rangle$ and parameterize it in terms of two form factors $D$ and $D_1$
\begin{eqnarray}
& & \!\!\!\!\!\!\!\!\!\!\!\!
\langle g^2 \phi(0,x) F_{\mu\nu}(x) \phi(x,0) F_{\lambda\rho}(0) \rangle \equiv  
\nonumber\\
& & \!\!\!\!\!\!\!\!\!\!\!\!
(\delta_{\mu\lambda}\delta_{\nu\rho} \! - \!\delta_{\mu\rho}\delta_{\nu\lambda}) D(x^2) 
\! + \! {1\over 2} \! \left[ {\partial\over\partial x_\mu}\! \left( x_\lambda \delta_{\nu\rho} 
- x_\rho \delta_{\nu\lambda} \right) \right.
\nonumber \\ 
& & \!\!\!\!\!\!\!\!\!\!\!\!
+ \left.  {\partial\over\partial x_\nu}\!
\left( x_\rho \delta_{\mu\lambda} - x_\lambda \delta_{\mu\rho} \right) \right] D_1 (x^2). 
\label{decom}
\end{eqnarray}
All the dynamics is contained in the form factors. The function $D$ is responsible for the area law 
and confinement. Both functions $D$ and $D_1$ are proportional 
to the gluon condensate $G_2 \equiv \langle \alpha_{\rm s} F^2(0) \rangle / \pi$.

In Ref. \cite{bali} we have examined a set of lattice data for the l.h.s. of Eq.(\ref{FF}).
We have found that: 1) it exists a window (roughly $x$ bigger than $0.2 \div 0.3$ fm) 
in which Lorentz invariance is restored in the continuum limit and therefore the r.h.s. 
of Eq. (\ref{FF}) is well approximated  by the two-point field strength correlator; 
2) in this window the functions $D$ and $D_1$ display an exponential 
decreasing behaviour $\simeq G_2 \exp\{- \vert x\vert / T_g\}$ with a correlation length 
$T_g \simeq 0.12 \div 0.19$ fm. For the details we refer to Ref. \cite{bali}.
These results compare well with other lattice determinations of these quantities which have 
been obtained by use of the so-called cooling method \cite{lat}.  However this is the first 
direct measurement without cooling and the first lattice evidence in favor of the Gaussian dominance 
in the Wilson loop average, 
\begin{equation} 
\langle W(\Gamma)\rangle \simeq e^{ -{1 \over 2} \int_{S(\Gamma)} \!\!dS_{\mu\nu}
\int_{S(\Gamma)} \!\!dS_{\rho\sigma}\, \langle g^2 \phi \, F_{\mu\nu} \phi \, F_{\lambda\rho} \rangle}
\label{wilab}
\end{equation}
In the long distance region (bigger than 0.2 fm) the Wilson loop
average is well approximated by the two-point field strength correlator and therefore 
the QCD infrared dynamics is ``effectively Abelian''. Moreover, the nonperturbative 
regime is controlled by two parameters: the gluon condensate $\langle F^2(0)\rangle$ 
and the gluon correlation length $T_g$. The  static quark-antiquark potential
is then given by \cite{DoSi}
$$
V_0(R) =  {1\over 2} \int_{|x_1|<R} \!\!\!\!\!\! d^2x ~(R-|x_1|)~D(x^2) 
+ {|x_1|\over 2}D_1(x^2)
$$
with  $d^2x = dx_1 dx_4$, $x^2 = x_1^2 + x_4^2$. In the limit $r/T_g \gg 1$, $V_0(r) \to \sigma r$ 
with the string tension given by
\begin{equation}
\sigma = {1 \over 2} \int d^2x  \,D(x^2).
\label{sigma}
\end{equation}
A non-vanishing $D$ function leads to confinement.

Quite naturally now a question arises. What is the relation between the ``Abelian behaviour'' 
of the Wilson loop average discussed above and the effective dual Abelian Higgs model 
suggested by the infrared lattice data mentioned in Sec. 1?

\section{DUAL EFFECTIVE DYNAMICS}

In Sec. 1 we have reported that the infrared dynamics of QCD seems to be effectively described by 
a dual Abelian Higgs model with action
\begin{eqnarray}
& & \!\!\!\!\!\!\!\!\!\!
S(C_\mu,\phi) = \int d^4x \left[ {1\over 4} G_{\mu\nu}(x) G_{\mu\nu}(x) \right.
\nonumber \\
& & \!\!\!\!\!\!\!\!
+ \left. {1\over 2}(D_\mu \phi)^*(x) (D_\mu \phi)(x) + V(\phi^*(x) \phi(x)) \right], 
\label{action}
\end{eqnarray}
where $G_{\mu\nu}(x) = \partial_\mu C_\nu(x) - \partial_\nu C_\mu(x) + G^s_{\mu\nu}(x)$,
$C_\mu$ being the dual field and $\displaystyle G^s_{\mu\nu}= g \, \varepsilon_{\mu\nu\alpha\beta}
\int_0^1 d\tau \int_0^1 d\sigma {\partial y_\alpha \over \partial \sigma} {\partial y_\beta
\over \partial \tau} \delta^4(x-y(\tau,\sigma))$  the Dirac field strength containing 
the quark currents flowing around the Wilson loop. $\phi$ is the Higgs field and 
$V(\phi^*\phi) = \displaystyle{\lambda \over 4} (\phi^*\phi -\phi_0^2)^2$ is the Higgs potential.  
The Higgs field is coupled to the gauge field $C_\mu$ via the covariant derivative 
$D_\mu \phi = (\partial_\mu + i e C_\mu)  \phi$.  The parameters of the model are the dual coupling 
constant $e= 2 \pi / g$, the Higgs coupling constant $\lambda$ and the Higgs vacuum expectation 
value $\phi_0$. The classical equations of motion follow from the action (\ref{action}). 
We emphasize that in this dual description the classical approximation does apply.

We have now two effective Abelian descriptions of the  Wilson loop average: 
Eqs. (\ref{decom})-(\ref{wilab}) and Eq. (\ref{action}).  
These are discussed in Ref. \cite{dual}. There the two-point field strength 
correlator (\ref{decom}) is related to the corresponding dual quantity 
${\cal G}_{\sigma\gamma\lambda\rho}(x,y) \equiv(\delta_{\lambda\sigma}\delta_{\rho\gamma} 
- \delta_{\lambda\gamma}\delta_{\rho\sigma}) \delta^4(x-y) 
- \epsilon_{\mu\nu\lambda\rho}\epsilon_{\beta\alpha\sigma\gamma} 
\partial^y_\beta \partial^x_\mu {\cal K}_{\nu\alpha}(x,y)$, where ${\cal K}_{\mu\nu}$ is the 
dual field propagator, and therefore to the underlying dual superconductor mechanism. 
We summarize here the main results.

\subsection{London limit without quark sources}

Let us consider the pure vacuum without quark sources. 
In the case $\lambda \to \infty$ we have $\phi=\phi_0$ and the Higgs field part of the 
action decouples. The dual field  propagator 
${\cal K}_{\mu\nu} \equiv \langle C_\mu C_\nu \rangle$, satisfies the equation: 
$$
(\partial^2 \delta_{\nu\mu} - \partial_\nu \partial_\mu 
-   e^2 \phi_0^2 \delta_{\nu\mu}) {\cal K}_{\nu\alpha}(x,y) \! = \! - \delta_{\mu\alpha}\delta^4(x-y) 
$$
with solution ($\infty$ stands for London limit) 
\begin{eqnarray}
{\cal K}^\infty_{\mu\alpha}(x,y) &=& \left( \delta_{\mu\alpha} 
- {\partial_\mu\partial_\alpha \over M^2} \right){\cal K}^\infty 
\nonumber\\ 
{\cal K}^\infty (x,y) &=& {M\over (2\pi)^2}{ K_1(M x) \over x}
\nonumber
\end{eqnarray}
with $M=e \phi_0$ the mass of the dual gluon. As a consequence we obtain 
\begin{eqnarray}
& & \!\!\!\!\!\!\!\!\!\!\!\!
{\cal G}_{\sigma\gamma\lambda\rho}(x)  \!=\!  (\delta_{\lambda\sigma}\delta_{\rho\gamma} - 
\delta_{\lambda\gamma}\delta_{\rho\sigma}) D^\infty(x^2)  + {1\over 2}\left[
{\partial\over\partial x_\lambda}\right. \nonumber\\
& & \!\!\!\!\!\!\!\!\!\!\!\! \left. (x_\sigma \delta_{\rho\gamma}  
- x_\gamma \delta_{\rho\sigma} )  
 +  {\partial\over\partial x_\rho} ( x_\gamma 
\delta_{\lambda\sigma} - x_\sigma \delta_{\lambda\gamma} ) \right]\! D^\infty_1(x^2)
\nonumber
\end{eqnarray}
with $D^\infty(x^2)  = \displaystyle {M^3\over 4\pi^2} {K_1(M x)\over x}$, 
$D^\infty_1(x^2) = -4 \displaystyle {{\rm d}\over {\rm d}x^2} K^\infty(x^2)$.
Therefore, assuming  the London limit of a dual Abelian Higgs action as a model of infrared QCD, 
the predicted form of the two-point field strength correlator as in Eq. (\ref{decom}), manifests 
a {\it non-vanishing} function $D$ and an exponential fall off for the functions $D$, $D_1$.  
Since ${\cal K}^\infty(x^2) {\mathop{\longrightarrow}\limits_{|x| \to \infty}} 
\displaystyle{1\over 2}{1\over (2\pi)^{3/2}}{1\over \sqrt{M} x^{3/2}} e^{-Mx} + \cdots$, 
the dual gluon mass $M$ plays the role of a correlation length. 

The logarithmic divergence of the string tension calculated via (\ref{sigma}) 
shows that the London limit is unphysical near quark sources where the Higgs 
field cannot be considered freezed to his vacuum value. In a physical situation 
we have to consider quarks and a Higgs field which vanishes on the Dirac string.

\subsection{Dual dynamics with quark sources}

Let us identify in the long range
$$
W(\Gamma) \simeq \langle e^{-S(C_\mu,\phi)} \rangle, 
$$
with the action $S$ given by Eq. (\ref{action}) and containing quark sources via the Dirac string. 
Using the classical equations of  motion we find 
\begin{eqnarray}
& & \!\!\!\!\!\!\!\!\!\! S(C_\mu^{\rm cl},\phi^{\rm cl}) = 
{1\over 2} \int_{S(\Gamma)} \!\! d S_{\sigma\gamma}(v) \int_{S(\Gamma)} \!\! d S_{\lambda\rho}(u) 
\nonumber\\
& & \!\!\!\!\!\!\!\!\!\! 
\times {\cal G}_{\sigma\gamma\lambda\rho}(v,u) 
 + \int d^4x \, \left[{1\over 2}(\partial \phi(x))^2 + V(\phi^2(x))\right]. 
\nonumber
\end{eqnarray}
However, neglecting the dependence of the Higgs field, via the equations of motion, 
on the strings, taking into account the contribution coming from the Higgs 
part as a finite contribution to the string tension, we find that, also in the general case, 
${\cal G}_{\sigma\gamma\lambda\rho}$ is equivalent to the QCD two-point non-local condensate 
(\ref{decom}) and in principle gives information on the validity of the decomposition (\ref{decom}) 
and on the existence and the behaviour of the functions $D$ and $D_1$. In the approximation in which 
the Higgs field depends only on the transverse coordinates, ${\cal G}$ depends only on the 
parallel components. In this approximation we can evaluate analytically the electric 
component of ${\cal G}$ and find that its behaviour is controlled by a form factor 
$D(x) = \displaystyle {1\over 2 \pi \ell^2} K_0\left({|x|\over \ell}\right)
{\mathop {\to}\limits_{ x\to \infty}} \sqrt{{\pi\over 2} {\ell \over |x|}} e^{-|x|/\ell}$
where $\ell = \sqrt{S_c}/S_\phi$, and
$S_c = \displaystyle{\mathop{\lim}\limits_{x_\perp \to 0}}
e \,C^{\rm np}(x_\perp)/x_\perp$ and $S_\phi = \displaystyle{\mathop{\lim}\limits_{x_\perp \to 0}}
e \phi(x_\perp)/x_\perp$ are constants known in terms of the Ginzburg--Landau parameter 
$\lambda/ e^2$. This suggests the identification of the correlation length $T_g$  with the dual 
quantity $\ell$. Notice that at variance with respect to the London limit result, here 
the correlation length is not simply given by the mass $M$ of the dual gluon.  

The static potential can be calculated exactly without the use of an ultraviolet cut-off  
\begin{eqnarray}
& & \!\!\!\!\!\!\!\!\!\!\!
V_0(R) = - {g^2 \over 4\pi}{1\over R} + {g^2 \over 2\pi} S_c \int_0^R \!\!\! dx_1 2 (R-x_1)
\nonumber \\
& & \!\!\!\!\!\!\!\!\!\!\!  \times \int_{-\infty}^{+\infty} \!\!\! dx_4 
{1\over 2\pi\ell^2} K_0\left({\sqrt{x_4^2+x_1^2}\over \ell}\right) 
+ \, {\rm Higgs} \, {\rm contr.}
\nonumber\\
& & \!\!\!\!\!\!\!\!\!\!\! =
 R \, {g^2 \over 2 \pi}S_c  + \left(e^{-R/\ell}-1\right){g^2 \over 2 \pi}S_c \,\ell 
- {g^2 \over 4\pi}{1\over R} + R \,\sigma_{\rm H} 
\nonumber \\
& & \!\!\!\!\!\!\!\!\!\!\! {\mathop{\longrightarrow}\limits_{ R \to \infty}}  R\,{g^2 \over 2 \pi}S_c 
+ R \,\sigma_{\rm H}, 
\label{string}
\end{eqnarray}
being the Higgs contribution to the static potential given by a linear term with string tension 
$\sigma_{\rm H}$. Taking explicitly into account this contribution, the total string 
tension is $\sigma= \displaystyle{g^2 \over 2 \pi}S_c+ \sigma_{\rm H}$. In particular,  
for a superconductor on the border ($\lambda/e^2 = 1/2$) we have $V_0(R)=  \pi \phi_0^2  \left( R + 
\ell\left(e^{-R/\ell}-1\right) \right) - \displaystyle{g^2 \over 4\pi}{1\over R}$. It is also 
possible to relate the Higgs condensate to the gluon condensate $G_2\simeq \lambda \phi_0^4$.

\section{CONCLUSIONS}

In  the assumption that the infrared behaviour of QCD is described by an effective 
Abelian Higgs model we have related the nonperturbative behaviour of the two-point field strength 
correlator $\langle g^2 F_{\mu\nu}(x)\, \phi(x,y)F_{\lambda\rho}(y)\phi(y,x)\rangle$ 
in QCD with the dual field propagator in the effective Abelian Higgs model of infrared QCD. 
In this way the origin of the non-local gluon condensate is traced back to an underlying 
Meissner effect and the phenomenological relevance of the Gaussian approximation 
on the Wilson loop is understood as following from the classical approximation in the dual 
theory of long distance QCD. In particular the correlation length $T_g$ of QCD, which we know 
from direct lattice measurements, can be expressed completely in terms of the dual theory 
parameters. We have calculated analytically the static potential and the string tension. 
It turns out that the string tension is given by an integral over a function of the correlation 
length which can be identified with the non-local gluon condensate. There is no 
cut-off introduced in this calculation since it is not performed in the London limit.

\end{document}